# GALACTIC PUNCTUATED EQUILIBRIUM: HOW TO UNDERMINE CARTER'S ANTHROPIC ARGUMENT IN ASTROBIOLOGY

Milan M. Ćirković
*Astronomical Observatory, Volgina 7,*
*11160 Belgrade, Serbia*
*e-mail:* mcirkovic@aob.bg.ac.yu

Branislav Vukotić
*Astronomical Observatory, Volgina 7,*
*11160 Belgrade, Serbia*
*e-mail:* bvukotic@aob.bg.ac.yu

Ivana Dragićević
*Faculty of Biology, University of Belgrade, Studentski trg 3,*
*11000 Belgrade, Serbia*
*e-mail:* dragicev@ibiss.bg.ac.yu

**Abstract.** We investigate a new strategy which can defeat the (in)famous Carter's "anthropic" argument against extraterrestrial life and intelligence. In contrast to those already considered by Wilson, Livio, and others, the present approach is based on relaxing hidden uniformitarian assumptions, considering instead a dynamical succession of evolutionary regimes governed by both global (Galaxy-wide) and local (planet- or planetary system-limited) regulation mechanisms. This is in accordance with recent developments in both astrophysics and evolutionary biology. Notably, our increased understanding of the nature of supernovae and gamma-ray bursts, as well as of strong coupling between the Solar System and the Galaxy on one hand, and the theories of "punctuated equilibria" of Eldredge and Gould and "macroevolutionary regimes" of Jablonski, Valentine, et al. on the other, are in full accordance with the regulation-mechanism picture. The application of this particular strategy highlights the limits of application of Carter's argument, and indicates that in the real universe its applicability conditions are not satisfied. We conclude that drawing far-reaching conclusions about the scarcity of extraterrestrial intelligence and the prospects of our efforts to detect it on the basis of this argument is unwarranted.

> An unflinching determination to take the whole evidence into account is the only method of preservation against the fluctuating extremes of fashionable opinion.
>
> Alfred North Whitehead, *Science and the Modern World* (1929), p. 268

> Thunderbolt steers all things.
>
> Heraclitus of Ephesus, *On Nature*, fragment B64

## 1. Introduction: Carter's argument

The well-known argument against the existence of extraterrestrial intelligence (henceforth ETI) due to the astrophysicist Brandon Carter (1983), and developed by various authors (e.g., Barrow and Tipler 1986), goes as follows. If astrophysical ($t_*$) and biological ($t_b$) timescales are truly uncorrelated, life in general and intelligent life in particular forms at random epoch with respect to the characteristic timescale of its astrophysical environment (notably, the Main-Sequence lifetime of the considered star). In the Solar system, $t_* \approx t_b$, within the factor of two. However, in general, it should be either $t_b \gg t_*$ or $t_b \sim t_*$ or $t_* \gg t_b$. The second case is much less probable *a priori* in light of the independent nature of these quantities. Carter dismisses the third option also, since in that case it is difficult to understand why the very first inhabited planetary system (that is, the Solar System) exhibits $t_* \approx t_b$ behaviour. On the contrary, we would then expect that life (and intelligence) arose on Earth, and probably at other places in the Solar System, much earlier than they in fact did. This gives us probabilistic reason to believe that $t_b \gg t_*$ (in which case the observation selection effect explains very well why we do perceive the $t_* \approx t_b$ case in the Solar System). Thus, the extraterrestrial life and intelligence have to be very rare, which is the reason why we have not observed them so far, in spite of the conjecture that favorable conditions for them exist at many places throughout the Galaxy.[1]

Carter's argument (henceforth CA) has been given much attention in the astrobiological and SETI-related discourses. It has been given very favorable assessment in the monograph of Barrow and Tipler (1986). Authors so diverse as Maddox (1984), Bostrom (2002), Barrow (2002), Lineweaver and Davis (2002, 2003), Davies (2003), and Duric and Field (2003) hold CA in high regard. Astrobiologists McKay (1996) and Livio (1999) regard it as an all-important constraint on astrobiological theories (although Livio presents an empirical counter-argument, he does not challenge the validity of the reasoning behind it). SETI skeptics, like Tipler (1988, 2004) use it as an important part of argumentation for claiming that we are alone in the Galaxy. Lineweaver and Davis (2002, 2003) mention CA approvingly, noticing that (i) it applies to noogenesis (origin of intelligence) rather than biogenesis (origin of life); and (ii) that any finding suggestive of long future duration of the biosphere will undermine it.

---

[1] Strictly speaking, this is just a half of Carter's argument in his fascinating 1983. paper. The rest concerns the issue of the number of "crucial" (or "critical") steps in emergence of intelligent observers on Earth and possibly elsewhere. Although later commentators, like Barrow and Tipler (1986) or Wilson (1994), devoted much attention to this "anthropic" prediction, it lies outside of the scope of the present paper. Without going into details, it is straightforward to see how undermining of the former argument leads to a substantial erosion of the claims of the latter.

However, it has also been criticized. Two most pertinent criticisms so far are those of Wilson (1994) and Livio (1999), the former mainly from the logico-methodological and the latter from the physical point of view. We shall use elements of their criticisms here, but the bulk of the present remarks have not been presented in the literature so far. Wilson's criticism is mostly methodological; he argues that Carter is wrong in restricting the range of possible relations of $t_b$ and $t_*$ to the three cases listed above, as well as that on the face of the argument, the fact that we appeared on Earth significantly before the end of the Sun's Main Sequence lifetime is highly unlikely. In addition, Wilson points out that the role of the anthropic reasoning in CA is very minor, almost trivial. In several places, Wilson vaguely alludes to some of the empirical inadequacies of CA (see below), but refrains from investigate them further; in the next sections, we shall try to perform that task.

The crucial assumption of CA is that there is no *a priori* reason for correlation between $t_*$ and $t_b$. Livio (1999, 2005) has pointed out that this is the main weakness of this argument; processes which *induce* correlations between the two timescales, like the oxygenation of the atmosphere on terrestrial planets, undermine the argument. Notably, if stellar UV radiation prevents the appearance of land life due to high absorption by nucleic acids and proteins, than it is critical that sufficient ozone layer is built before land life appears. This, in turn, might induce a correlation between the astrophysical and biological timescales, since various stellar masses (and thus various lifetimes) will generate different amounts of UV radiation and dictate various rates of oxygenization of the atmospheres of hypothetical planets in their respective habitable zones. With the fall of this essential assumption of independence of two timescales, CA is doomed as well.

We would like to hereby express even more radical criticism of Carter's argument, based on the questioning of its basic premises, and to show that the reasoning behind it is inherently flawed—at least so without additional assumptions of rather questionable validity. The main idea is quite simple: Carter's argument relies on the assumption that there are *fixed* (or at least well-defined) and *roughly known* timescales for at least astrophysical processes. In addition, it is assumed that the relevant biological timescale is well-defined, albeit unknown, as well. We reject these assumptions, and intend to show that there is sufficient physical justification to propose alternatives. These alternatives are more complicated than Carter's simplification, but this is the necessary price to pay to be in accordance with tremendous achievements of modern astrophysics and astrobiology during the last decade or so. (In addition, Carter requires that the relevant timescales are *independent*, but this was criticized by Livio and others, and is only a part of the present argument.) These alternatives encompass (i) external physical forcings acting on local biospheres all over the Galaxy, and (ii) elements of complex evolution, namely quasi-periodicity, stochasticity, change of macroevolutionary regimes, and secular evolution with cosmological time. In other words, the core element of CA, belief that "the probability of intelligence increases monotonically with time" (Barrow and Tipler 1986; see §3 below) is (A) just a case of special pleading, and (B) is likely to be wrong on empirical grounds.

In a very limited form, this central argument has been sketched in Dragićević and Ćirković (2003). As Einstein memorably used to say: "Everything should be made as simple as possible, *but not simpler*." [emphasis by the present authors] In particular, we believe that CA violates the second part of this important methodological guideline by failing to take into account timescale correlations induced by both secular evolution of the Galaxy and sudden catastrophic events; this stands in full accordance with Whitehead's maxim quoted above, when applied to recent and current debates on SETI and related projects. In addition, it is a microcosm of several traditional issues in philosophy of science in general, and philosophy of biology in particular: issues of inevitability vs. contingence, gradualism vs. catastrophism, local vs. global influences on the biosphere, position of intelligent observers on the "tree of life", and some others, recur in our study of CA and related topics.

It is sometimes stated that CA offers an example of the scientific nature of the anthropic reasoning, by virtue of its falsifiability: it offers a prediction that our current astrobiological and SETI efforts will fail and that we shall not discover extraterrestrial intelligent beings in the Milky Way. Formally speaking, this does indeed make the hypothesis scientific, but it can be argued that in this case the meaning of falsifiability is stretched beyond its reasonable usage. For instance, the statement "There are no intelligent alien species in the Galaxy" presumes our capacity to always discriminate between intelligent and non-intelligent *aliens* with certainty, which can hardly be taken for granted (cf. Raup 1992; Lem 1976, 1987). Even if it were, the timescales for this kind of falsification are quite outstanding; in fact they are *at least* equal to the often-cited Fermi-Hart timescale for visiting (or colonizing) all stars in the Milky Way.[2] Even the most ardent Popperian should pause when faced with such a remote prospect of falsification, especially when it is not necessary to doubt the anthropic reasoning itself in order to contest a specific argument using many auxiliary assumptions.

It is important to emphasize that we do not intend to make a case for the existence of ETIs in the Milky Way. That is a quite distinct (and, arguably, much more formidable) task. Our aim is simply to show how a particular anti-ETI argument—strengthened, unfortunately, by endless uncritical repetitions in both research and popular literature—can be undermined. Only insofar our lack of credence in the existence of Galactic ETIs is based on CA, it can be said that our study offers an indirect support for ETI *plausibility*. There are, however, other anti-ETI arguments—notably the Tsiolkovsky-Fermi-Viewing-Hart-Tipler's argument, usually known simply as Fermi's paradox (for the best reviews see Brin 1983; Webb 2002)—which are beyond the scope of the present study and which could, in principle, support ETI skepticism even if CA is dismantled. In our view, the entire problem of the existence or absence of extraterrestrial life and intelligence remains completely open.

---

[2] The qualification is necessary given our inability to decisively determine, for example, the degree of intelligence and consciousness of some of the species we share our planet with, namely marine mammals (e.g. Browne 2004). Since it is easier to make oneself known than to establish or refute intelligence on an alien planet with certainty, expected timescale can only be larger than the Fermi-Hart limit (cf. Ćirković and Bradbury 2006).

**2. Are there well-defined timescales?**

There are many cases in everyday life, as well as in science, where apparently independent quantities are of similar or even the same order of magnitude. In an amusing example in his classic textbook Peebles (1993, p. 366) jovially notes the coincidence between the Eddington limit[3] on luminosity of a star per unit mass, and Peebles' own luminosity per unit mass. Although we are today vitually certain that the cosmic microwave background is of primordial origin, this does not invalidate the famous coincidence noted by Sir Fred Hoyle (1994; see also Lightman and Brawer 1992) that the quantity of helium in the universe is almost exactly such that its synthesis through the fusion of hydrogen *in toto* will produce about the same amount of energy as contained in CMB photons. The timescale for reading of this paper is of the same order of magnitude as the variability timescale of the ultraluminous X-ray source M74 X-1 (and, indeed, most of the microquasars in the local universe!). It would be preposterous and epistemologically naive to assume that in each instance of similar timescales, phenomena have to be causally linked. Such coincidences are so ubiquitous in our complex universe that entire pseudo-sciences have long ago arisen around some of them (measurements of the Great Pyramid of Egypt, for instance); they are reflection of humanity's psychological need for finding causal links and explanations even where all there is are, in G. Udny Yule's famous term, "nonsense correlations". By far most of correlations in the world are of this, noncausal nature. In Peebles' words, "for practical purposes it is only an accident of essentially unrelated numbers."

But CA is an example of exactly the opposite extreme: denial of the possibility of such coincidences actually occuring, contrary to what both history of science and our everyday experiences tell us. According to Carter, even if they are observed to occur, as is allegedly the case in the Solar System, this must not reflect anything deeper than a consequence of our restricted viewpoint. Thus, Carter ignores some sound advice of Agatha Christie's famous Miss Marple of: "Any coincidence is always worth noticing... you can throw it away later if it is *only* a coincidence." In effect, CA is spectacularly underestimating the vast complexity and intricacy of Nature.

Even if there is no causal link between $t_b$ and $t_*$, it would be erroneous to reject the $t_b \sim t_*$ case as Carter does. How many orders of magnitude does this region possess? Are there different *external* constraints on the timescales, precluding them having values in the entire (0, +∞) range? This has also been criticized by Wilson (1994), though without appeal to physical reality, which makes the case against Carter's thesis significantly stronger. We shall argue now that, due to the oversimplification, there are additional timescales, which make $t_b \sim t_*$ the most interesting case. Then, it becomes an additional benefit that such choice would make the Earth truly unexceptional and thus is in good agreement with the Copernican principle.

---

[3] The Eddington limit is set by the balance of gravity and radiative pressure by stellar radiation in the star's envelope (for a fine review, see Kippenhahn and Weigert 1994). If the envelope is to be bound by the star, the radiation pressure force has to be balanced by the gravity. Thus, the Eddington limit bounds the stellar luminosity and explains, among other things, why stars similar to the Sun but billion times brighter do not exist.

Let us first redefine the astrophysical timescale $t_*$ as the timescale of *continuous habitability* of a terrestrial planet in the Milky Way galaxy. The difference may sound pedantic, but is in fact crucial when we recognize that (astro)physical processes *other than the evolution of its parent star* can influence the habitability of a planet. In particular, the need to abandon the "closed box" astrobiological picture of Earth (and terrestrial planets in general) is emphasized in a number of recent studies from different points of view. Most pertinently, Lineweaver et al. (2004) investigate the concept of the Galactic Habitable Zone (henceforth GHZ), introduced by Gonzalez et al. (2001), comprising the stars in the Milky Way potentially possessing habitable planets with complex life (for a fine review, see Gonzalez 2005).[4] In both astrobiology and the Earth sciences, such paradigm shift toward an interconnected, complex view of our planet has already been present for quite some time in both empirical and theoretical work (e.g., Clube and Napier 1990; Cockell 1998; Burgess and Zuber 2000; Lenton and von Bloh 2001; Franck et al. 2000, 2001; Carslaw et al. 2002; Iyudin 2002; Gies and Helsel 2005; Chyba and Hand 2005).

(It is important to understand here that the very talk about habitable zones makes the assumptions of independent events at best suspicious. Habitable zones are *defined* as spatio-temporal regions where conditions for life arise due to *correlated* processes. As far as prospects for SETI are concerned, the relevant zone is GHZ, which occurs as a consequence of roughly understood processes of chemical and dynamical evolution of the Milky Way and its stellar populations. Even more telling in this respect is the concept of the Cosmic Habitable Age (CHA), introduced by Gonzalez (2005). Insofar as habitable zones are an unavoidable part of the modern astrobiological discourse, any argument based on the independent development of biospheres automatically looses force.)[5]

Before we analyze the particulars of these external influences and consequent timecale forcing, we wish to emphasize that the very idea of Carter that Main Sequence stellar lifetimes are the only relevant (astro)physical timescales is already a dangerous simplification. There are some rather uncontroversial—in contrast to some of the ideas considered below in more

---

[4] As kindly pointed out to us by Prof. David Grinspoon, the first suggestion of anything even remotely similar to GHZ was given by the great author and philosopher Stanislaw Lem in his *One Human Minute* (Lem 1986). Lem obviously foreshadowed and inspired much of the contemporary research in astrobiology, including the present study (esp. Lem 1987).

[5] A minor additional argument to the same effect may come from the panspermia hypotheses which, although quite speculative and uncertain, have experienced recent resurgence. Thus, Napier (2004), as well as Wallis and Wickramasinghe (2004) have constructed working panspermia models in agreement with all known astrophysics.

detail—counterexamples in both past and future history of our planet.[6] For instance, the "Snowball Earth" episodes occurring at least twice in the geological past (Kirschvink et al. 2000; Hoffman et al. 1998) represent global catastrophes which may have annihilated all life except for the small habitats around marine volcanoes and hydrothermal vents. It is entirely plausible that similar episodes of severe global glaciation could have annihilated all life at Earth-analogs elsewhere, so that the "astrobiological clock" gets a complete reset possibly even without any external causative agent, but due to an unfortunate combination of movement of continental plates and Milankovich cycles.[7] Similarly, it seems clear that geophysical processes governing the carbon-silicate cycle are sustainable for a time shorter than the Main Sequence timescales in at least a fraction of potentially inhabitable terrestrial planets in the Milky Way (e.g., Lindsay and Brasier 2002; Gerstell and Yung 2003; Ward and Brownlee 2002). This was not known at the time of Carter's 1983 article. Any such large-scale trends make CA *a posteriori* less appealing, since they induce further correlations and have their own quasi-deterministic timescales, thus undermining the independence assumption.

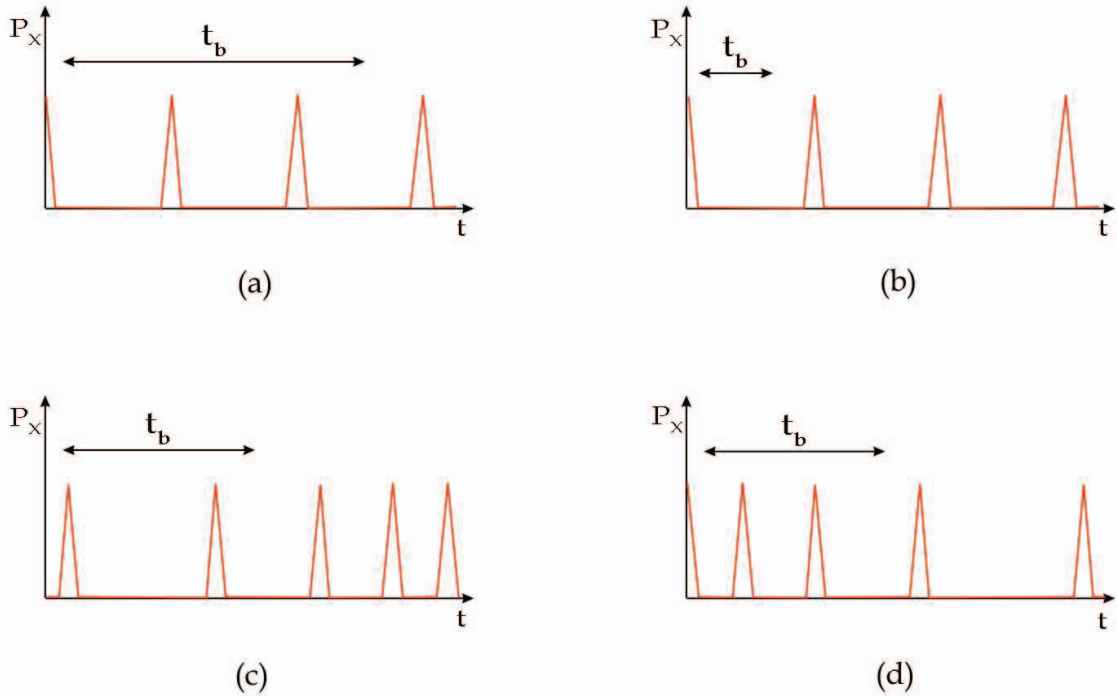

**Figure 1.** Schematical presentation of possible relationship of two independent timescales. With $t_b$ we denote the median of biological timescales on different planets of GHZ. We may assume that $P_X$ corresponds to some measure of the extinction probability in the most general sense. While cases (a) and (b) correspond to the situation envisaged by Carter's rendition of the anthropic reasoning (in particular, case (a) is the situation encapsulated by CA), we argue that these situations are unjustified simplifications.

---

[6] Parenthetically, this refutes the claim that we have even a *single data point* of quite unambiguous meaning. This, in our view, underlines the hubris of those who use this (uncertain) point to grandeloquently conclude that we are the only intelligent species in the Galaxy.
[7] However, for a view ascribing even the "Snowball" glaciations to our astrophysical environment, see Pavlov et al. (2005).

Physical reality corresponds to cases *at least* as complex as cases (c) and (d), where we have the environment monotonously becoming either more hostile or friendlier to life. In these cases, obviously, we have to take into account another timescale, which describes the rate of increase or decrease of the extinction events.

This is related to the important issue of biotic feed-back. Another consequence of discoveries in Earth sciences and astrobiology in the last decade or so is that the existence of life on Earth tends to make it more habitable both complexity- and time-wise. Simple lifeforms induce changes in the environment conductive to the appearance of more complex lifeforms, and, even more pertinently from the present point of view, the existence of both simple and complex life tends to increase the timespan of the habitable Earth beyond the bounds set by the Main Sequence evolution of the Sun (Lenton and von Bloh 2001). This not only sheds some new light on the controversial Gaia hypothesis, but also shows that the probability of observing an inhabited planet within a given sample at a particular time is not a linear function of the probability of biogenesis, as one would naively expect (and what is assumed in CA). This does not represent an argument against CA yet, since the latter takes the total lifetime of a star on the Main Sequence as the ultimate limit on the timespan of the biosphere, which remains true irrespective of the feedback.[8] However, it does much to weaken the spirit, if not the letter of CA, since it shows that probability of finding life at a particular place cannot be a linear function of time *ceteris paribus*. (We shall return to this point below.)

We do not need to emphasize that the biological timescales are still very poorly understood. There are some claims that the timescale of biological evolution on Earth is fairly typical. Russell (1983, 1995) claims that the appearance of intelligent beings occurs on the average about 3 Gyr after the initial stages of planetary formation. Much shorter timescales have been proposed: McKay argues that plate tectonics actually *delayed* the appearance of complex lifeforms on Earth, by keeping the level of oxygen low for a long time. According to that idea, the duration of the Precambrian could be as low as $10^8$ yrs on planets without plate tectonics, such as Mars (McKay 1996). This would, in turn, significantly accelerate the emergence of sufficient complexity as a precondition for intelligence.

It is in this astrobiologal key that we can reiterate part of Wilson's (1994) criticism of CA contained in the following passage:

> At first glance, the claim that $t_e$ should not differ from a given value of $\bar{t}$ seems to be equivalent to the claim that $\bar{t}$ should not differ from a given value of $t_e$. But these claims are fundamentally different. The reason the latter one is invalid is that $\bar{t}$, insofar as it represents the time that evolution is intrinsically most likely to require, is a probabilistic or statistical quantity. Our knowledge of the value of such quantity cannot be significantly enhanced by the evidence of a single case, especially the nonrandomly chosen one of our own evolution. Only if we were to become aware of a large number of actual cases of extraterrestrial evolution and their corresponding timescales, or if we were to advance our knowledge of the timescales governing various evolutionary mechanisms, could we provide a reasonable estimate of $\bar{t}$, and perhaps eliminate values of $\bar{t}$ much less than $\tau_0$. But given only $t_e \sim \tau_0$, we cannot on the basis of this single evidential sample conclude much at all about $\bar{t}$. We certainly cannot eliminate, as Carter thinks we can, the possibility that $\bar{t} \ll \tau_0$.

---

[8] If we discard intentional action of intelligent lifeforms, which we do throughout this study, as the subject is still too speculative; it seems intuitively clear that advanced technological civilizations would be able to extend the lifetime of a biosphere significantly beyond the natural limits, using technologies such as stellar lifting (Criswell 1985).

## 3. Hidden uniformitarianism

Consider the different situations described in Figure 1. With obvious simplification, we imagine extinction probability of lifeforms as generally very low, except for short "spikes" which may correspond either to recurring (similar to mass extinction episodes in Earth's history) or single adversary events (for instance, the end of stellar evolution), i.e. everything which is subsumed in Carter's astrophysical timescale $t_*$. Now, if there is a well-defined biological timescale, CA can be represented as choosing between the cases shown in (a) and (b). Clearly, CA suggests that we should accept case (a), in which the probability of appearance of life on an average terrestrial planet in the Galaxy is minuscule. However, what about cases (c) and (d)? It is clear that in these cases the governing timescale is the one associated with the increase or decrease in frequency of the extinction spikes. It follows that
   (1) "Carter's criterion" of the relationship between the biological and astrophysical timescales is time-dependent and not universal; and
   (2) that we may need additional timescales, linked to all astrophysical processes which can cut or impede biological evolution.
The problem is that, in order to accept a picture like (c) or (d) we need to abandon one of the most cherished prejudices of the nineteenth and most of the twentieth century, which is uniformitarianism of rate (or gradualism). It is not only that the habitability of an astrobiological site is not constant in time, but the frequency of important events ("extinction spikes") is changing with the evolution of the Galactic system. We shall argue below that the situation shown in Fig. 1.d is the best model for astrobiology and that in such framework CA fails. As we have said, there are a host of recent indications that the Solar System is in fact an open system, strongly interacting with its Galactic environment (e.g., Rampino and Stothers 1985; Rampino 1997; Leitch and Vasisht 1998; Shaviv 2002; Melott et al. 2004; Pavlov et al. 2005; Gies and Helsel 2005). Interactions induce correlations; correlations ruin arguments based on independence assumptions and coincidences. Why is that simple fact so widely shunned in favor of a prejudice representing essentially a return to the outdated nineteenth-century Lyellian gradualism?

Barrow and Tipler (1986) succintly state this critical uniformitarian assumption for which we need a specific label:

**THESIS (*):** "the probability of intelligence <u>increases monotonically</u> with time" (p. 559, our emphasis)

We deny this assumption, for reasons to be discussed below. It is important to understand that (*) is the central plank of CA—with it gone, the whole edifice crumbles.

First, (*) encapsulates an anthropocentrism unwarranted even in the simplest local case of the biological evolution on Earth. It is by no mean clear (and many evolutionary biologists have denied it) that the history of the terrestrial biosphere represents anything even remotely describable as "monotonical" approach to intelligence. Even if we take the most merciful interpretation of "monotonically", which allows for the ever-present paleontological stasis, there is simply no indication in the history of life that intelligence is inherently more probable today than, say, in the middle of the Cretaceous or in $10^8$ years from now. (Quite contrary to the seeming intention of Barrow and Tipler, it is anthropic reasoning which tells us that we should *not* invoke our presence to argue for the thesis (*)—it is trivial observation that our discussing the subject matter shows that intelligence exists now, while it does not tell us anything about its intrinsic probability. Moreover, the same fact suggests that the intelligence posing these questions is not very old, at least compared to geological or astrophysical timescales, since it is hardly conceivable that any significantly older intelligent species would not have much better insight into the nature of intelligence. Anthropic reasoning, when properly applied, is not just disteleological, but actively anti-teleological; see, e.g. Bostrom 2002.)

Second, the problem is not that the probability of intelligence increases with time *ceteris paribus*. It is quite clear that the probability of observing any particular physically possible phenomenon at least once increases with cosmic time; the very statistical nature of our world ensures that. Of course, we need to take into account our cosmological knowledge: if the lifetime of our universe is finite – as was believed in now mostly discredited recollapsing "Big Crunch" models – then most physically allowed configurations of matter will simply have no time to arise accidentally due to statistical fluctuations. Contrariwise, if the time is infinite and the world is finite and stationary on large scales (as was commonly thought in the time of Boltzmann, say), than any configuration of matter in accordance with the general conservation laws will be achieved countless times, no matter how *a priori* improbable.[9]

But this is entirely different from the claim that we have a monotonic "ascent" towards intelligence under very specific (and in the cosmological context very atypical) conditions required for biogenesis and biological evolution. Such a monotonic approach entails some specific causal reason, since both the spatial and temporal timescales we are considering here are many *hundreds* of orders of magnitude smaller than those required for the random assembly of even the simplest living systems. However, such a causative agent has not been found, and is most likely to join the outdated (vitalism) and/or compromised new-age notions (morphogenetic field).[10] In Stephen Jay Gould's (1984) words: "the failure to find a clear 'vector of progress' in life's history... [is] the most puzzling fact of the fossil record." And if that is true for a single, by astrobiological measure, physically stable and uniform terrestrial biosphere, we have grounds for accepting it *a fortiori* for the set of (actual or potential) biospheres comprising GHZ.

---

[9] On exactly such grounds Boltzmann and his assistant Shuetz argued that the present low-entropy state of the universe is just a fluctuation within a much larger reality which is perpetually in the state of heat death. This picture is now, of course, outdated, but can be given a modern reformulation; see Ćirković (2003).

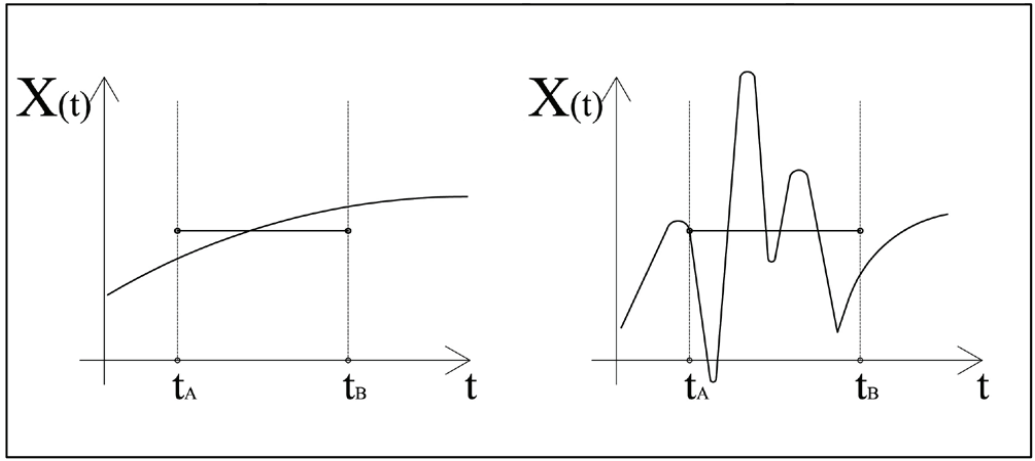

**Figure 2.** Is mean really the message? It is a notorious truism that the timescale of measurement dictates the amount of information we can get about evolving phenomena: the longer our measurement lasts, the less useful information we get due to averaging inherent in any measurement. In the context of CA we can regard noogenesis as an extremely slow type of "measurement": thus, it is likely that the physical conditions all over GHZ vary substantially on shorter timescales, which precludes getting information on the basis of a single astrophysical timescale.

Therefore, the thesis (*) represents a particularly illuminating example of what biologists came to call a chain-of-being fallacy: the quasi-Victorian idea that the history of biosphere is a steady, linear progression through more and more complex forms culminating in a whiskey-sipping, golf-playing, white-clad and well-armed gentlemen (occassionally subjugating uncouth barbarians overseas). As shown by Gould in the very first chapter of *Wonderful Life*, this iconography has been conventionally employed in support of various scientifically wrong, but socially comforting, ideological social issues (Gould 1989). Although fierce debates on the issue of "progress"—or large-scale evolutionary trends in general—continue to this day (e.g., Dawkins 1989; Dennett 1995; Gould 1996, 2002; Shanahan 1999, 2001; Knoll and Bambach 2000; Carroll 2001; Conway Morris 1998, 2003), both sides do agree that the chain-of-being picture is untenable. No serious biologist will today defend the idea that humankind is the pre-determined pinnacle of Nature. (Sadly, this realignment has not been followed by the popular press, especially the part with its own ideological axe to grind.) Thus, (*) is ideologically loaded: it supports the idea that intelligence is fundamentally different from other biological traits, and that the bearers of intelligence are entitled to higher and more important place in the natural order of things.

---

[10] This indeterminism should not be confused with the erroneous view that biological (or at least Darwinian) evolution proceeds randomly in the metaphysical sense. The (in)famous "Boeing 747" argument of Sir Fred Hoyle has been often cited and misused by creationists and other pseudo-scientists as the "proof" of intelligent design or some similar ideologically concocted scheme. Fortunately enough, the argument is demonstrably wrong; it has been refuted many times, notably by Dawkins (1989) and Dennett (1995).

Ironically, distinguished biologists who have *opposed* SETI like Mayr and Simpson, devoted a large part of their professional careers to debunking the chain-of-being fallacy! Notably, the adaptationist paradigm of which Mayr is one of founding fathers, even hesitates to ascribe any particular importance to intelligence, or to proclaim it different from any other trait in nature. Within the framework of adaptationism, there is no a priori difference between intelligence and, say, the spiral form of the shell of *Nautilus*. Now, just imagine rephrasing (*) in the following form:

> **THESIS (*''):** "the probability of a spiral shell with a pitch angle between 23° and 25° increases monotonically with time".

In our view, (*'') is almost obvious nonsense; why should we then—if we discard sentimentality, anthropocentrism and possible extrascientific agendas—give better treatment to (*)? If, to use Stephen Jay Gould's famous metaphor, "the tape was rewound" to the time of the Cambrian Explosion, it would be highly unlikely for humans to re-appear after sufficient time has elapsed (Gould 1989). Gould forcefully argued in several books and papers (Gould 1985, 1987, 1989, 1996), that the very notion of "progress" of the terrestrial biosphere is highly suspicious, culture-laden and with very slim empirical support (if at all). How much more pretentious and vacuous does it sound when applied to the immense diversity that other Galactic environments may present! If the thesis (*) is an interpretation of "progress" or "ascent", the same classical criticisms apply.

McKay (1996) playfully suggests that dinosaurs could in fact have developed not only intelligence, but spacefaring civilization as well, without our noticing it at present! Strong erosion, following the giant impact, in the course of 65 Myr could easily obliterate all traces of such a predilluvian culture. Similarly, if humans were to go extinct soon (perhaps as a result of runaway climatic catastrophe, nuclear winter or a misuse of biotechnology or nanotechnology[11]), in a few million years all traces of human civilization would have been obliterated (except for the satellites in stable orbits and a couple of long-range space probes). Would the next intelligent species, if it ever arises subsequently, have the same cultural predilection for (*) as we have?

Now, if we conclude that (*) is unjustified for the simplest local model of biological evolution on a single planet, how likely is that (*) will apply to a large set of habitable planets comprising GHZ? Even if the planets were isolated, "closed-box" idealizations—which is unrealistic—the uniform behavior implied by (*) is as probable as sudden motion of molecules of homogeneous air in a room into a 1 m$^3$ volume in a corner. As Boltzmann, Zermelo, Culverwell, and others already knew in the nineteenth century, such conspiratorial behavior is highly unlikely on statistical grounds *without going into detailed physics* of thermodynamical systems (e.g., Steckline 1983). *Per analogiam*, without knowing any details on particular astrobiological development in each habitat, we might argue that uniformity of evolution expressed by (\*) is improbable. Of course, our discussion here and elsewhere pertains only to the evolution of hypothetical biospheres comprising GHZ, which are of interest to SETI studies; if we take into account other galaxies, clusters, etc. the situation may be both quantitatively and qualitatively different.

---

[11] An option pertinent for obvious reasons—and very relevant for contemporary astrobiologists (e.g., Bostrom and Ćirković 2007).

When we reject hidden uniformitarianism even the $t_b \ll t_*$ case of Carter's dilemma is not to be rejected so lightly. An obvious counterexample in this respect is the much-debated "impact frustration" of early lifeforms (Raup and Valentine 1983; Maher and Stevenson 1988; Oberbeck and Fogleman 1989). It is entirely conceivable that early terrestrial life appeared independently several times, only to be destroyed by catastrophic impacts during the epoch of the so-called late heavy bombardment. Only when the frequency of impacts decreased sufficiently (perhaps after the end of the late-heavy bombardment), were early lifeforms capable of spreading, diversifying and evolving in order to produce the subsequent rich and complex terrestrial biosphere. If this was so, $t_b' \simeq 10^7 - 10^8$ yr $\ll t_*$ where with $t_b'$ we denote the timescale for biogenesis; while we cannot still infer anything about "true" $t_b$ (i.e. the noogenesis timescale), if for any reason there were an upper limit to $t_b'$ (related perhaps to the chemical evolution of Earth's atmosphere or surface and the increase of Solar radiation flux), it could be perfectly conceivable that life could "miss the last train" due to the impact interruptions. This scenario is highly instructive, since it shows (A) strong coupling between lifeforms and physical environment, and (B) timescale forcing, through which a physical timescale (the interval between major impacts) actually becomes the only relevant quantity. While we cannot treat this topic in detail here, this example of a new development in astrobiology bolsters the conclusion that uniformitarianism and continuous habitability of a planet are just convenient and oversimplifying myths.

**4. A plausible alternative: global extinction mechanisms**

If we accept that CA is at least severely limited by the non-uniformitarian history of life it is natural to ask for more details about the major non-uniformities which impede its "monotonic" progress.[12] Fortunately, modern astrophysical research offers much in this respect. An important paper of Annis (1999) opened a new vista by introducing (though not quite explicitly) the notion of a *global regulation mechanism*, that is, a dynamical process preventing or impeding uniform emergence and development of life all over the Galaxy.[13] In Annis' model, which he dubbed the phase-transition model for reasons to be explained shortly, the role of such global Galactic regulation is played by gamma-ray bursts (henceforth GRBs[14]), collosal explosions caused either by terminal collapse of supermassive objects ("hypernovae") or mergers of binary neutron stars. GRBs observed since the 1950s have been known for more than a decade to be of cosmological origin. Astrobiological and ecological consequences of GRBs and related phenomena have been investigated recently in several studies (Thorsett 1995; Dar 1997; Scalo and Wheeler 2002; Thomas et al. 2005). To give just a flavor of the results, let us mention that Dar (1997) has calculated that the terminal collapse of the famous supermassive object Eta Carinae could deposit in the upper atmosphere of Earth energy equivalent to the simultaneous explosions of 1 kiloton nuclear bomb per $km^2$ all over the hemisphere facing the hypernova! According to the calculations of Scalo and Wheeler (2002), a Galactic GRB can be lethal for eukaryotes up to the huge distance of 14 kpc. Thus, this "zone of lethality" for advanced lifeforms is bound to comprise the entire GHZ whenever a GRB occurs within the inner 10 kpc of the Galaxy. Annis suggested that GRBs could cause mass extinctions of life all over the Galaxy (or GHZ), preventing or arresting the emergence of complex life forms. Thus, there is only a very small probability that a particular planetary biosphere could evolve intelligent beings in our past. However, since the regulation mechanism exhibits *secular evolution*, with the rate of catastrophic events decreasing with time, at some point the astrobiological evolution of the Galaxy will experience a change of regime. When the rate of catastrophic events is high, there is a sort of quasi-equilibrium state between the natural tendency of life to spread, diversify, and complexify, and the rate of destruction and extinctions. When the rate becomes lower than some threshold value, intelligent and space-faring species can arise in the interval between the two extinctions and make themselves immune (presumably through technological means) to further extinctions.

---

[12] We do not presume any special understanding of evolutionary biology here; the formulation is applicable to a case of stable, only very slowly changing biological environment—a single macroevolutionary regime in terms of Jablonski (1986, 1989)—in which Barrow and Tipler's metaphor of "monotonic approach to intelligence" might perhaps work. Such world is not the real world, but its juxtaposition with the real world can teach us some important lessons.

It is important to understand that the GRB-mechanism is just one of several possible physical processes for "resetting astrobiological clocks". Any catastrophic mechanism operating (1) on sufficiently large scales, and (2) exhibiting secular evolution can play a similar role. There is no dearth of such mechanisms; some of the bolder ideas proposed in the literature are cometary impact-causing "Galactic tides" (Asher et al. 1994; Rampino 1997), neutrino irradiation (Collar 1996), clumpy cold dark matter (Abbas and Abbas 1998), or climate changes induced by spiral-arm crossings (Leitch and Vasisht 1998; Shaviv 2002). Moreover, all these effects are cumulative: *the total risk function of the global regulation is the sum of all risk functions of individual catastrophic mechanisms*. The secular evolution of all these determine collectively whether and when conditions for the astrobiological phase transition of the Galaxy will be satisfied. Of course, if GRBs are the most important physical mechanism of extinction, as Annis suggested, then their distribution function will dominate the global risk function and force the phase transition.

---

[13] A similar suggestion has been made earlier by Clarke (1981), although his model was entirely qualitative and used wrong physical mechanism (Galactic core outbursts) for global regulation; see also Clube (1978), LaViolette (1987).

[14] The literature on GRB physics is voluminous; some of the best review articles are Piran (2000), Dar and De Rújula (2004).

One should also note that there is another sort of global regulation mechanism which we shall not discuss here, but which features prominently in the SETI-related discourse: an intentional global regulation imposed by whichever intelligent community first achieves Kardashev's Type III status (utilizing resources on the pan-Galactic basis). This is the astrobiological background of scenarios such as the "Zoo hypothesis" (Ball 1973) or the "Interdict hypothesis" (Fogg 1987). The same regulatory effect could be achieved without directly controlling the Galactic resources by releasing destructive von Neumann probes (see the disturbing comments on this scenario in the classical review of Brin 1983). It is very difficult to assess the real value of such scenarios while lacking deeper theoretical ideas about the capacity of advanced technological civilizations (possibly *postbiological* in nature; see Dick 2003; Ćirković and Bradbury 2006). It seems reasonable to assume a gradual switch of regimes from natural to artificial astrobiological regulation. All these ideas present *special cases* of the general global regulation hypothesis, but are too speculative to be seriously considered at present.

GRB regulation has an important correlation property: the rhythm of biological extinctions should be *synchronized* (up to the timescales of transport times ~$10^4$ yrs for γ-rays and high-energy cosmic rays) in at least part of the histories of all potentially habitable planets. In fact, a bold hypothesis has been put forward recently by Melott et al. (2004) that a known terrestrial mass extinction episode, one of the "Big Five" (the late-Ordovician extinction, cca. 440 Myr before present), corresponds to a Galactic GRB event.

It is intuitively clear that such correlated behavior undermines Carter's argument. With a set of modest additional assumptions it is possible to show it quantitatively. For instance, in Figure 3 we show results of toy numerical experiments performed in order to see how timescale forcing arises in simplified evolving systems. This presents a simple realization of the astrobiological regulation model of Annis (1999). GRBs are taken to be random events occuring with exponentially decreasing frequency

$$\nu(t) = \nu_0 \exp\left(-\frac{t}{t_\gamma}\right),$$

with the fixed characteristic timescale $t_\gamma = 5\,\text{Gyr}$ in accordance with the cosmological observations (e.g., Bromm and Loeb 2002), and biological timescales for noogenesis are randomly sampled from a log-uniform distribution between $10^8$ (the minimum suggested by McKay 1996) and $10^{16}$ yrs (the total lifetime of the Galaxy as a well defined entity; Adams & Laughlin 1997). For simplicity it has been assumed that the age of the Galaxy is exactly 12 Gyr and that all planets are of the same age. It is taken that the chain of events leading to life and intelligence can be cut by a catastrophic event at any planet in our toy-model Galaxy with probability $Q$, and its astrobiological clock reset. The toy model counts only planets achieving noogenesis *at least once* and it does not take into account any *subsequent* destructive processes, either natural or intelligence-caused (like nuclear or biotech self-destruction). Probability $Q$ can, in the first approximation, be regarded as a geometrical probability of an average habitable planet being in the "lethal zone" of a GRB, and more complex effects dealing with the physics and ecology of the extinction mechanism can be subsumed in it.

While we shall present more detailed analysis and interpretation of these and similar numerical experiments in a forthcoming work (Vukotić and Ćirković 2007), some conclusions invariably support our criticism of CA and are worth mentioning here. The system exhibits a systematic shift of behavior as we move from small values of $Q$ (gradualism) to large values (catastrophism). At large $Q$, we have a step-like succession of astrobiological regimes, governed by external timescale forcing. In each regime, it is obvious that the ages of inhabited planets are not independent and uncorrelated, just the contrary, as we expected from the considerations above.

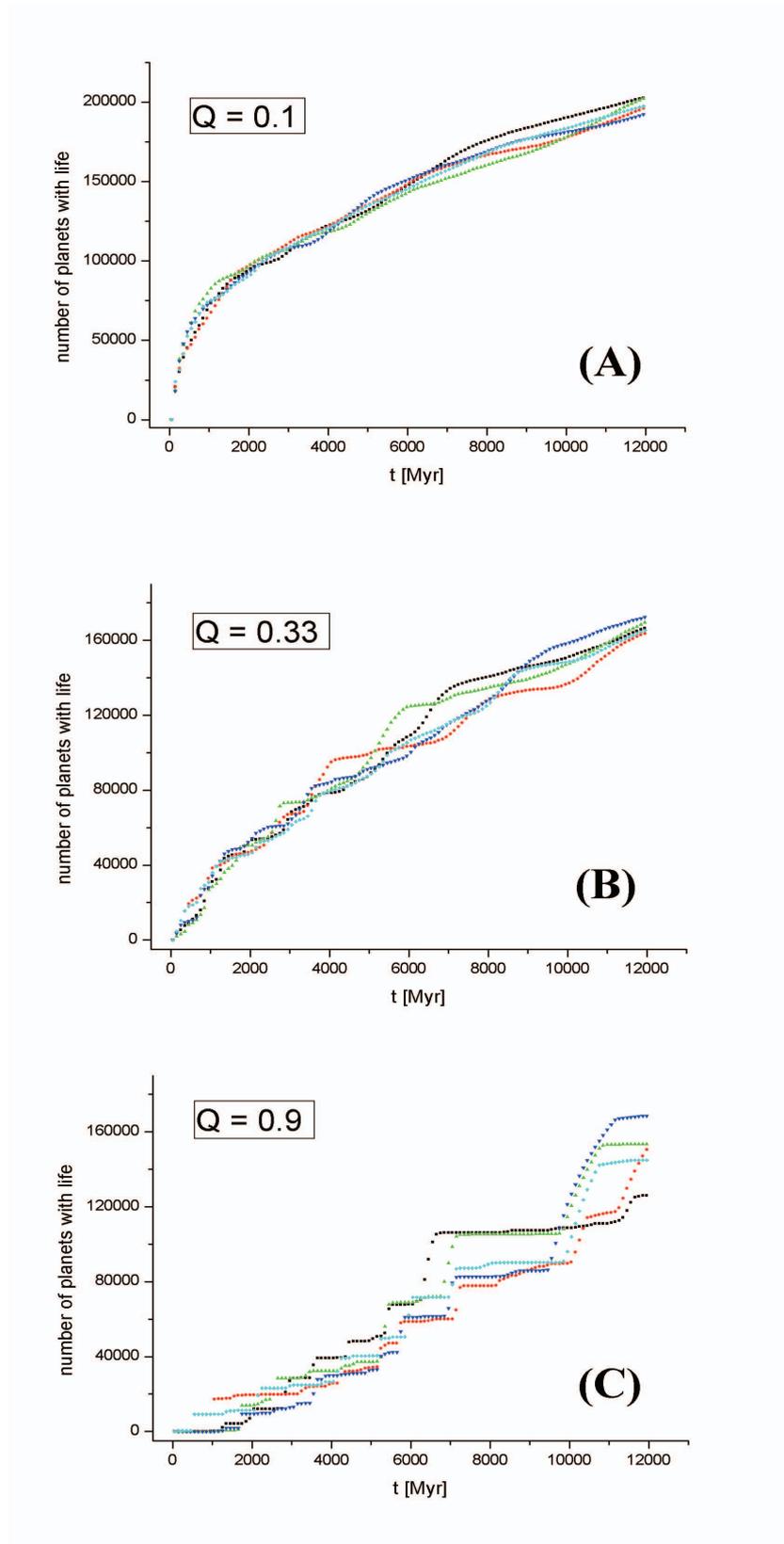

**Figure 3.** Results of a simple Monte Carlo simulation demonstrating timescale forcing in the toy model of a galaxy with N = 10⁶ habitable planets. Three cases of different average extinction probability *Q* are given (with 5 color-coded runs for each case), where small *Q*

represents an approximation to the classical uniformitarian case, while $Q \to 1$ corresponds to catastrophe-driven evolution.

In other words, neocatastrophism removes, ironically enough, the basic tacit assumption of CA. If the agents of extinction are correlated over the spatial scale of GHZ, timescale forcing undermines Carter's reasoning in a natural way. As Heraclitus fancied 25 centuries ago, (astrophysical) thunderbolt may indeed steer all things (astrobiological).

(It is important to emphasize that in the simulation above we have neglected all sources of correlations between the life-bearing sites barring the GRB regulation. Some processes – like panspermia, either natural or directed – certainly deserve to be taken into account, and we shall discuss them in detail in a forthcoming study. However, all these processes will only strengthen the correlations and thus decrease our confidence in Carter's reasoning.)

## 6. Conclusions

We conclude that it is too early to draw sceptical conclusions about the abundance of extraterrestrial life and intelligence from our single data point via the "anthropic" argument of Carter (1983). In addition to other deficiencies of the argument pointed out in the literature, we emphasize that a picture in which regulation mechanisms reset local astrobiological clocks (which, consequently, tick rather unevenly) offers a way to reconcile our astrophysical knowledge with the idea of multiple habitats of life and intelligence in the Galaxy. In other words, Earth may be *rare in time, not in space*! Quite contrary to the conventional wisdom, we should not be surprised if we encounter many "Earths" throughout the Galaxy at this particular moment in time, at stages of evolution of their biospheres similar to the one reached at Earth. The unsupported assumption of gradualism is identified as the main source of confusion and unwarranted SETI skepticism (for a related discussion in the practical context of the Drake equation see Ćirković 2004b). This pertains to the Milky Way galaxy, where communication times are short enough to make the entire effort worthwhile (and to bring other factors, such as Fermi's paradox, into play). If we take into account progressively larger ensembles it will be possible sooner or later to find the monotonic behaviour criticized above, but this is largely formal and irrelevant for practical SETI.

The astrobiological picture presented here can be understood by means of loose analogy with the much-discussed theory of punctuated equilibrium in evolutionary biology. Seeking to explain the evident stop-start nature of the fossil record Eldredge and Gould (1972) proposed the theory of punctuated equilibria (for the detailed elaboration and synthetic view see Gould 2002). According to this theory, species tend to remain stable for long periods of time ("stasis"). The equilibrium is punctuated by abrupt changes in which existing species are suddenly (on geological timescale) replaced. The astrobiological analogy of paleontological stasis can be found in Fig. 3c, where we perceive long periods (of ~ 1 Gyr duration) with the same number of inhabited planets before a sudden change. This feature is in itself antithetical to the spirit of CA; as emphasized by Gould (1989):

> Hence, a good deal more than half the history of life is a story of prokaryotic cells alone, and only the last one-sixth of life's time on earth has included multicellular animals. Such delays and long lead times strongly suggest contingency and a vast realm of unrealized possibilities. If prokaryotes had to advance toward eukaryotic complexity, they certainly took their time about it.

This is directly opposite to the thesis (*) and its monotonic ascent toward a (perceived) noble goal.

This is intimately linked to the issue of existence or otherwise of a well-defined biological timescale, universal for all habitable planets in the Milky Way. Obviously, in order to discuss this issue we have first to establish to which degree observed timescales on Earth (the one for appearance of life, or the one for rise of complex metazoans, or the one for emergence of intelligent species) are a consequence of deterministic or just contingent processes, and how big a role chance has played in their values (Carroll 2001). Gould's "paradox of the first tier" points in the same direction: "…mass extinctions are sufficiently frequent, intense, and different in impact to undo and reset any pattern that might accumulate during normal times." (Gould 1985) Here we add another aspect to this "enlightened" view of catastrophes: not only do they provide the pump of evolution by enabling innovative overthrows of entire faunas, but – in the astrobiological context – they could provide us with correlations on the basis of which we could meaningfully consider astrobiological evolution of the Galaxy; indirectly, they offer weak support for our current and future SETI efforts.

This undermining of Carter's argument is entirely in accordance with the well-known tendency in the history of science and the human culture in general: overcoming of the sense of privilege surrounding the Solar System, Earth, terrestrial life, and humanity. This "Copernican" tendency comes as a *consequence* of the astrophysical discourse, not as a sacred dogma to be preserved at all costs. Furthermore, a natural generalization of the history of the terrestrial biosphere to the case of the Milky Way from the astrobiological point of view entails the acceptance of a sort of Galactic (neo)catastrophism (or a Galactic punctuated equilibrium!). It immediately undermines CA, since there is no fixed, unique, reified timescale in the core of the argument.

There are several reasons of a partly non-scientific nature for the strong impression CA leaves in many quarters. As we have seen, it is tempting to subsume all complicated astrophysics and planetary science into a single timescale, although this degree of simplification is, in fact, unwarranted; this applies to biology even more forcefully. There is an unfortunate tendency in the philosophy of science to downplay radical scientific theories and underestimate our present level of understanding; sometimes it is motivated by healthy reasons of skepticism, but often—the case of Mach's fierce "philosophical" opposition to Boltzmann's atomism comes to mind—it actually represents a conservative backlash, impeding recognition of new ideas. In addition, the textbook account of the defeat of catastrophism and the misleading philosophical legacy imparted to it in the mid-nineteenth century lead all too often to half-conscious neglect of any temporal markers in investigation of natural phenomena other than the beginning and the end. Finally, CA offers an emotionally satisfying, but nevertheless false sense of "strength in numbers".

A particularly dangerous form of "quick and dirty" generalization is embodied in the thesis (*) of monotonical ascent toward intelligence criticized above. Naive chain-of-being anthropocentrism (or intelligence-centrism) surrounding the reasoning of CA-proponents is starkly manifested here. In our view, *neither* is there a physical basis (uniformitarianism of environmental conditions) for it, *nor* there is clear biological justification (since the adaptive value of intelligence is still an unknown quantity). Thus, CA represents more wishful thinking coupled with intellectual inertia when faced with abandoning gradualism and the closed-box assumption, than a serious scientific argument.

The tremendous progress in astrobiology (e.g. Chyba and Hand 2005) clearly demonstrates that the oversimplifications inherent in CA are no longer tenable. To retain them means to reject all we have achieved in the last couple of decades on establishing concrete physical and chemical conditions for emergence of life in the cosmic context. If CA is largely, as Carter himself admitted, an argument from ignorance, then *any decrease in our ignorance ought to prompt its reassessment and reevaluation*. We are fortunate enough to live in this exciting epoch of truly wonderful results in this field, from cosmology and orbital observatories down to biochemical labs and paleontology museums. That it is also the epoch in which CA can be effectively undermined is by no means a *coincidence*.

All in all, we have no *a priori* (or even anthropic-based) reason to reject the existence of extraterrestrial intelligence in the Milky Way. Geocentrism stays defeated and the road for serious SETI studies is as open as ever.

**Acknowledgements.** Our foremost thanks go to David Grinspoon, for his kind help, encouragement, and exceptionally detailed and insightful comments on a previous version of this manuscript. We also wish to thank Branislav Nikolić, Robert J. Bradbury, Tanja Berić, Nick Bostrom, Srdjan Samurović, Milan Bogosavljević, Irena Diklić, Vesna Milošević-Zdjelar, Anders Sandberg, Richard Cathcart, Mark A. Walker, Dejan Urošević, James Hughes, Saša Nedeljković, Nikola Milutinović, Dušan Inđić and Marija Karanikić for useful discussions and/or valuable technical help. The Editor and the two referees for *Astrobiology* are also hereby acknowledged, as well as are Paul Farrell of Copernicus Books and Simon Mitton of the Cambridge University Press. This research has made extensive use of NASA's Astrophysics Data System Abstract Service, as well as of the services of the KOBSON consortium of libraries. M. M. Ć. and B. V. have been partially supported by the Ministry of Science of the Republic of Serbia through the project ON146012, "Gaseous and stellar components of galaxies: interaction and evolution".